\documentclass[12pt]{article}
\usepackage{epsfig}

\usepackage{graphicx}
\usepackage{cancel}
\usepackage{amssymb,amsmath}

\def\harr#1#2{\smash{\mathop{\hbox to .3in{\rightarrowfill}}
 \limits^{\scriptstyle#1}_{\scriptstyle#2}}}

\def\s2{\frac{1}{\sqrt2}}

\def\beqa{\begin{eqnarray}}
\def\eeqa{\end{eqnarray}}

\def\Dsl{\,\raise.15ex\hbox{/}\mkern-13.5mu D} 
\def\d3{d^3}

\newcommand{\im}{\mathrm{i}}

\newcommand{\be}{\begin{equation}}
\newcommand{\ee}{\end{equation}}
\newcommand{\beq}{\begin{eqnarray}}
\newcommand{\eeq}{\end{eqnarray}}

\usepackage{amssymb,amsmath}

\usepackage{mathrsfs}

\def\be{\begin{equation}}
\def\ee{\end{equation}}
\def\beqa{\begin{eqnarray}}
\def\eeqa{\end{eqnarray}}

\begin{document}

\begin{center}
\Large{\bf Euclidean Wormholes in Ho$\check{\rm r}$ava-Lifshitz
Gravity} \vspace{0.5cm}

\large  H. Garc\'{\i}a-Compe\'an\footnote{e-mail address: {\tt
compean@fis.cinvestav.mx}}, Alberto V\'azquez\footnote{e-mail
address: {\tt aivazquez@fis.cinvestav.mx}}

\vspace{0.3cm}

{\small \em Departamento de F\'{\i}sica, Centro de
Investigaci\'on y de Estudios Avanzados del IPN}\\
{\small\em P.O. Box 14-740, CP. 07000, M\'exico D.F., M\'exico}\\

\vspace*{1.5cm}
\end{center}

\begin{abstract}
We study Euclidean wormholes in the framework of the Ho$\check{\rm
r}$ava-Lifshitz theory of gravity. Euclidean wormholes first
appeared in the Euclidean path integral approach to quantum gravity.
In a more general way, Hawking and Page interpreted such
configurations as solutions to the Wheeler-DeWitt equation with
appropriate boundary conditions. We use the projectable version of
Ho$\check{\rm r}$ava-Lifshitz gravity to obtain the Wheeler-DeWitt
equation of a minisuperspace model considering a closed Friedmann
Universe plus a massless scalar field. For large values of the scale
factor we find that the solution of the Wheeler-DeWitt equation
coincides with the one obtained by Hawking. Whereas in the limit
corresponding to the early Universe we find a new set of solutions,
which agree with the Hawking and Page boundary conditions for
wormholes.

\vskip 1truecm

\end{abstract}

\bigskip

\newpage

\section{Introduction}
\label{sec:intro} It is well known that in the quantum field
theoretic description of Einstein's gravity one founds ultraviolet
(UV) divergences. The presence of these UV divergences has its
origin in the fact that Newton's constant has mass dimension
$\left[G_{N}\right]=-2$. That means that the gravitational
interaction may be described by an effective field theory, in which
case it would require an UV completion. In Ref. \cite{Horava:2009uw}
Ho$\check{\rm r}$ava proposed one possible UV completion of Einstein
theory. The basic idea of Ho$\check{\rm r}$ava's theory is to
improve the UV behavior of gravity by adding higher-order terms in
the spatial component of the curvature to the Einstein-Hilbert
action. The result is a theory of gravity in 3+1 dimensions that is
power-counting renormalizable and whose equations of motion are of
second order in time, avoiding the presence of ghosts.  In this
scenario Ho$\check{\rm r}$ava's idea was to follow some work of
Lifshitz in condensed matter physics regarding the consideration of
an anisotropic scaling of space and time
\begin{equation}\label{scaling}
t \rightarrow b^{z} t, \qquad x^{i} \rightarrow b x^{i}, \qquad
(i=1,2, \ldots, d)
\end{equation}
This scaling is known as Lifshitz scaling, where $b$ is a constant,
$d$ denotes the spatial dimension of spacetime, and $z$ is a number
called the dynamical critical exponent, which dictates the degree of
anisotropy between space and time. It is clear that Lorentz symmetry
is broken for $z \neq 1$ in the UV and it is recovered only when
$z=1$, which occurs at low energies in the infrared (IR).

It is important to point out that the Lifshitz scaling does not
respect the full invariance under diffeomorphisms present in general
relativity (GR). Thus one has to introduce a more restricted
reparametrization of spacetime which respect the foliation ${\cal
F}$ into space and time
\begin{equation}
t \mapsto \tilde{t}(t), \quad  x^{i} \mapsto \tilde{x}^{i}(t,x^i).
\end{equation}
This remaining redundancy is called
foliation-preserving-diffeomorphisms (or ${\cal F}$Diff for short)
because the spatial diffeomorphisms are the ones that remain
unchanged. In other words, it is chosen a preferred time.  This
proposal constitutes the so called Ho$\check{\rm r}$ava-Lifshitz
(HL) theory of gravity and it can be regarded as an extension of GR
to higher energies (for some reviews on the subject, see
\cite{Weinfurtner:2010hz,Sotiriou:2010wn,Wang:2017brl,Mukohyama:2010xz}
and references therein).

In this article we will work in one of the versions of the HL theory
known as the {\it projectable} theory. It is known that this theory
has a unphysical scalar degree of freedom which leads to a
perturbative IR instability \cite{Sotiriou:2010wn}. For the
projectable case this IR instability was studied in Refs.
\cite{Mukohyama:2010xz,Izumi:2011eh,Gumrukcuoglu:2011ef}. In these
references it is found that in the limit $\lambda \to 1$, the
resulting theory is GR coupled to the scalar model describing dark
matter (DM). In the specific case of quantum cosmology in the
minisuperspace approach for a homogeneous and isotropic metric of
the Fiedmann-Lemaitre-Robertson-Walker (FLRW) type, the projectable
model leads to a Friedmann equation including a DM component through
the additional scalar mode. From the view point of the canonical
approach to gravity, projectable and {\it non-projectable} models
differ in the local nature of the Hamiltonian constraint for the
non-projectable case and non-local one for the projectable case. In
the non-projectable model the scalar degree of freedom is absent.
Consequently, if one is only interested in solving the Hamiltonian
constraint (i.e. the Wheeler-DeWitt (WDW) equation) for a FLRW
metric, both: projectable and non-projectable models will give the
same results and the IR perturbative instability is not evident.
Some of the papers describing some solutions of Ho$\check{\rm
r}$ava-Lifshitz's gravity in quantum cosmology in the minisuperspace
are
\cite{Bertolami:2011,Christodoulakis:2011np,Pitelli:2012,Vakili:2013wc,Obregon:2012bt,Benedetti:2014dra,Cordero:2017egl}.

On the other hand, the idea of {\it wormhole} solution in GR was
suggested by Wheeler, the ideas is that the topology of spacetime
may fluctuate on scales of the order of the Planck length ($\sim
10^{-33}$ cm) giving rise to wormhole configurations
\cite{Wheeler:1955zz,Anderson:1986ww}. These objects are defined as
finite-action solutions to the equations of motion of the classical
Euclidean Einstein field equations, which is why they are also
called {\it gravitational instantons}. In other words, they are
Euclidean metrics that describe two asymptotically flat regions
joined by a narrow tube or throat. For an account on some of these
subjects in Euclidean gravity, see for instance,
\cite{Gibbons:1994cg}.

The interest on Euclidean wormhole physics peaked in the late 1980s
after Giddings and Strominger found a gravitational instanton
solution by considering a model of an axionic field (a 3-form)
coupled to gravity \cite{Giddings:1987cg}. The importance of this
solution was appreciated mainly due to the application of wormholes
to the cosmological constant problem. Such an idea was pursued by
Coleman who used the saddle point approximation in the path integral
approach to show that one of the possible effects of wormholes was
to set the value of the cosmological constant to zero
\cite{Coleman:1988tj}. Later Hawking also argued that macroscopic
wormholes might be responsible for the mechanism of black hole
evaporation. In this picture, baby universes are pinched-off from
some region of our Universe carrying away information
\cite{Hawking:1990jb}. Euclidean wormholes have plenty of
implications on particle physics and cosmology
\cite{Hebecker:2018ofv}, and for that reason a renewal interest has
started to appear in recent years. In particular, Euclidean
wormholes (or more specifically, Giddings-Strominger axionic
wormholes) seem to satisfy the Weak Gravity Conjecture
\cite{Hebecker:2016dsw}, which gives support to its physical
relevance.

For some time wormholes were only studied as solutions to the
Euclidean field equations using a semiclassical treatment. Such
solutions exist only for specific kinds of matter. Thus it seems
more natural that their importance rely in microscopic physics and
one shall to study them in a quantum mechanical setting. This
approach was followed by Hawking and Page in \cite{Hawking:1990in}.
They regarded wormholes as solutions to the WDW equation obeying the
so called {\it Hawking-Page boundary conditions}: $({\rm a})$ The
wave function needs to be exponentially damped for large
3-geometries; $({\rm b})$ It is regular as the 3-geometry collapses
to zero.

The foliation of spacetime in constant time hypersurfaces is known
as the Arnowitt, Deser and Misner (ADM) \cite{Arnowitt:1962hi})
decomposition of spacetime \cite{Arnowitt:1962hi}. It is also the
starting point of quantum cosmology, where the quantum dynamics is
governed by the WDW equation (for a review of quantum cosmology, see
for instance, \cite{Halliwell:1990uy,Wiltshire:1995vk}). The
formalism of quantum cosmology turns out to be an important tool to
investigate the implications of HL gravity at the quantum level. In
the present paper we study Euclidean wormholes in the context of
Ho$\check{\rm r}$ava-Lifshitz quantum cosmology. Some work on
wormhole solutions in the context of Ho$\check{\rm r}$ava-Lifshitz
gravity have been given in Refs.
\cite{BottaCantcheff:2009mp,Son:2010bs,Bellorin:2014qca,Bellorin:2015oja}.
In the present work, we are most concerned with obtaining solutions
to the WDW equation that satisfy the Hawking-Page conjecture for
Euclidean wormholes. Although solutions of this type have been found
in minisuperspace models for GR coupled to both massless and massive
scalar fields
\cite{Hawking:1990in,Ruz:2013hfa,SolomonsPhDthesis1993,Carlini:1995zi,Carlini:1997sy},
not much attention has been given to the canonical approach of
wormholes, commonly known as Euclidean quantum wormholes
\cite{Kim:1997fq}. For that reason, we will explore more that path,
considering  Ho$\check{\rm r}$ava-Lifshitz gravity, which as a
proposal of a UV completion of GR, seems to be more appropriated for
exploring the early Universe regime.

We will propose a minisuperspace model of quantum cosmology
starting, for simplicity, from the action of the projectable version
of HL gravity. However, in order to obtain wormhole solutions to the
WDW equation we need to couple matter to gravity. The theory of
scalar fields coupled to HL gravity, requires a theory of
higher-order spatial derivatives of the scalar discussed in Refs.
\cite{Kiritsis:2009sh,Calcagni:2009ar,Tawfik:2016dvd,Mukohyama:2009gg}.
Following this approach, we end up with a matter action that reduces
to a theory with minimal coupling in the IR. In fact, for a
homogeneous scalar field, the matter action behaves as the
relativistic one. However, in our analysis this will not be the case
because we will take the scalar field as a perturbation. Thus, the
idea is to couple a (massless) scalar field to HL gravity, and then
quantize the model to obtain the corresponding WDW equation. All of
this is done by considering the FLRW metric of a closed Universe.
Finally we will find solutions of the WDW equation that satisfy the
Hawking-Page conditions for quantum wormholes.

This article is organized as follows, in Section 2, we begin by
introducing the action of the projectable version of HL gravity as
well as the action of the scalar matter fields. In Section 3 we give
a brief review of the Hawking-Page conjecture for Euclidean quantum
wormholes and then outline one of their solutions found when
considering a model with conformally invariant matter. In Section 4,
we obtain the quantum cosmological model for the projectable HL
gravity coupled to a scalar field, which we treat as a perturbation.
We find explicit asymptotic solutions of the WDW equation in the
limit of the early and late Universe. For some of these solutions
such limits satisfy the Hawking-Page boundary conditions. In the
same Section we extend the analysis for cases of non-vanishing
negative and positive cosmological constant. We argue if these
solutions represent wormholes. Finally in Section 5 we give our
conclusions and final remarks.

\section{Ho$\check{\rm r}$ava-Lifshitz gravity}
\label{sec:REV}

We start from the Einstein-Hilbert action in its ADM form
\cite{Arnowitt:1962hi,Halliwell:1990uy,Wiltshire:1995vk}, i.e.
written in terms of the the 3-metric $h_{ij}$ of the spatial surface
$\Sigma$, and the extrinsic curvature
\begin{equation}
\label{extrinsicindex}
K_{ij}=\frac{1}{2N}(\dot{h}_{ij}-D_i N_j-D_jN_i),
\end{equation}
where $N$ is the lapse function and $N ^i$ is the shift vector.
Ho$\check{\rm r}$ava's theory modifies such action by adding
higher-order spatial curvature terms in order to obtain a
renormalizable theory of gravity in 3+1 dimensions.

We will use the projectable version of the theory, which takes the
lapse function to be dependent only on time, $N=N(t)$. The action is
given by \cite{Weinfurtner:2010hz,Sotiriou:2010wn,Wang:2017brl}
\begin{align}\label{horavagravityaction}
S_{\text{HL}}=\frac{M_{\mathrm{P}}^{2}}{2} \int  \mathrm{d} t \mathrm{d}^{3}x N \sqrt{h}\Big\{&K^{i j} K_{i j}-\lambda K^{2}-2 \Lambda +R+ M_{\mathrm{P}}^{-2}\left(g_{2} R^{2}+g_{3} R_{i j} R^{i j}\right) \notag \\
&+M_{\mathrm{P}}^{-4}\left(g_{4} R^{3}+g_{5} R (R_{i j} R^{i j})+g_{6} R_{j}^{i} R_{k}^{j} R_{i}^{k}\right) \notag \\
&+M_{\mathrm{P}}^{-4}\left[g_{7} R D^{2} R+g_{8}\left(D_{i} R_{j
k}\right)\left(D^{i} R^{j k}\right)\right]\Big\},
\end{align}
where the $g_n \, (n=0,\dots,8)$ are dimensionless running coupling
constants, $M_{\mathrm{P}}$ is the Planck mass, $D_i$ stand for
covariant derivatives associated to the metric $h_{ij}$ where $h$ is
its determinant, and $R_{ij}$, $R$ are the Ricci tensor and scalar
curvature of the spatial surface $\Sigma$, respectively.

The parameter $\lambda$ runs under the renormalization group flow
\cite{Mukohyama:2010xz}. In particular, in the IR limit, $\lambda
\to 1$, and all the higher-order curvature terms go to zero ($g_n
\to 0$ for $n=2, \dots, 8$). Therefore, in principle one would
recover GR. However, as we mentioned in the introduction Section and
in \cite{Cordero:2017egl}, this IR limit is unstable in the case of
the projectable theory. The perturbative analysis shows
\cite{Izumi:2011eh} that there is an scalar degree of freedom that
does not decouple from the gravitational field. Thus it is necessary
to perform a non-perturbative approach and to restore the GR limit
by non-linear dynamics. In the present situation we are studying the
dynamics at the level of the Hamiltonian constraint, not from the
full Einstein equations (Friedmann equations) point of view. Thus,
for our particular aim of the quest of Euclidean wormholes solutions
to the WDW equation for the projectable model, the instability will
not be evident and will not play a direct role in the description.

\subsection{Adding matter to the theory}
We write a total action of the following form
\begin{equation}\label{totalhl}
S=S_{\text{HL}}+S_{\text{m}},
\end{equation}
where $S_{\text{HL}}$ is the action of the projectable version of HL
gravity, and $S_{\text{m}}$ is the matter action. We will focus only
on the case of scalar matter. The action has to be compatible with
all the symmetries of the theory. The general action is that of a
non-relativistic scalar field, which has a quadratic kinetic term
and a superposition of terms with higher-order spatial derivatives
of the scalar field. We write the action as
\cite{Kiritsis:2009sh,Calcagni:2009ar,Tawfik:2016dvd,Mukohyama:2009gg}
\begin{equation}\label{scalarHL}
S_{\text{m}}=\frac{1}{2}\int dt d^{3} x\sqrt{h} \, N\left[\frac{(3
\lambda-1)}{2} \frac{(\dot{\phi}-N^{i} \partial_{i}
\phi)^2}{N^{2}}+F\left[\partial_{i} \phi, \phi\right]\right].
\end{equation}
This action is strongly motivated by the original Lifshitz scalar
theory. It obeys the Lifshitz scaling with dynamical critical
exponent $z=3$, and the ${\cal F}$Diff symmetry. The factor $(3
\lambda-1)$ is introduced for future convenience.

To have UV renormalizability, the function $F$, should contain up to
six spatial derivatives \cite{Fujimori:2015wda}. We write $F$ as
follows
\begin{equation}
F\left[\partial_{i} \phi, \phi\right]=c_1\Delta-c_2 \Delta^{2}+c_3
\Delta^{3}-V(\phi),
\end{equation}
where $\Delta=\partial_i \partial^i$ is the 3d Laplacian associated
to the metric $h_{ij}$, $V(\phi)$ is a potential term and $c_i$ are
constants which are related to the energy scale, that is
\begin{equation}
c_2 =\frac{1}{M^2},\quad c_3=\frac{1}{M^4}.
\end{equation}
The constant $c_1$ agrees with the velocity of propagation of light
in the IR, which in our units is set to one.

In the UV fixed point, the matter action is given by
\begin{equation}
S_{\text{m}}^{\text{UV}} \sim \frac{1}{2} \int dt d^{3} x \sqrt{h}
\, N\left[\frac{(3 \lambda-1)}{2} \frac{(\dot{\phi}-N^{i}
\partial_{i} \phi)^2}{N^{2}}-c_3 \phi \Delta^{3} \phi\right].
\end{equation}
In other words, the operator $\mathcal{O}=c_3 \phi \Delta^{3} \phi$,
dominates in the UV.

On the other hand, in the IR, Lorentz invariance is restored, and
when $\lambda \to 1$ we end up with a relativistic scalar matter
action with an arbitrary potential
\begin{equation}
S_{\text{m}}^{\text{IR}} \sim \frac{1}{2} \int d t d^{3} x \sqrt{h}
\, N\left[\frac{(3 \lambda-1)}{2} \frac{(\dot{\phi}-N^{i}
\partial_{i} \phi)^2}{N^{2}}-c_1\partial_{i} \phi \partial^{i} \phi
-V(\phi)\right].
\end{equation}

\section{The Hawking-Page quantum wormholes}
Before we proceed to introduce our model. We will briefly review the
so-called quantum wormholes of Hawking and Page
\cite{Hawking:1990in}. In the context of quantum cosmology, the
quantum wormholes are solutions to the WDW equation satisfying
certain boundary conditions. In contrast, classical wormholes are
Euclidean metrics (Wick rotated metrics, $t \to -\im \tau$) which
are solutions to the Euclidean classical field equations
representing spacetimes consisting of two asymptotically look-like
flat Euclidean regions joined by a narrow tube or throat.

We begin the quantum treatment of wormholes by introducing a
3-surface $\Sigma$, which is a cross-section of the wormhole that
separates two asymptotically Euclidean regions. We will also
consider matter fields $\phi$ on $\Sigma$. Then, we describe the
quantum state of the wormhole by the wave functional
$\Psi[h_{ij},\phi]$, where $h_{ij}$ is the 3-metric on $\Sigma$. The
wave function obeys the WDW equation
\begin{equation}
\left(-\frac{2\kappa}{\sqrt{h}}G_{ijkl}\frac{\delta^2}{\delta h_{ij}
\delta h_{kl}}-\frac{\sqrt{h}}{2 \kappa} \, (^{(3)}{R}{}-2\Lambda) +
\widehat{\mathcal{H}}^{\text{matter}}\left[\phi,\frac{\delta}{\delta
\phi}\right]\right)\Psi[h_{ij},\phi]=0,
\end{equation}
where $G_{ijkl}:={\sqrt{h} \over 2} (h_{ik}h_{jl} + h_{il}h_{jk}
-h_{ij}h_{kl})$ is the DeWitt metric, and $\kappa$ is Newton's
constant, see \cite{Halliwell:1990uy,Wiltshire:1995vk}.

One has to solve the WDW equation, and then impose certain boundary
conditions to obtain the quantum state of the wormhole. These
boundary conditions have to express the fact that the 4-metric is
non-singular, and has two asymptotically Euclidean regions. This is
difficult to implement in superspace, that is why one considers
minisuperspace models. Following
\cite{Hawking:1990jb,Hawking:1990be,Hawking:1990in,Hawking:1988ae},
we will work with the Euclidean Friedmann closed Universe plus a
small perturbation $\varepsilon_{ij}$
\begin{equation}
d s^{2}=N^{2}(\tau) d \tau^{2}+a^{2}(\tau)\left(\Omega_{i
j}+\varepsilon_{i j}\right) d x^{i} d x^{j},
\end{equation}
where $\Omega_{i j}$ is the 3-metric of a unit 3-sphere, ${\bf
S}^3$. To be more precise, we have chosen $\Sigma$, the
cross-section of the wormhole, to be the 3-sphere ${\bf S}^3$.
Hence, the quantum state $\Psi$ that we need to find is that of a
closed Friedmann Universe.

The 3-metric on ${\bf S}^3$ is then
\begin{equation}
h_{i j}=a^{2}\left(\Omega_{i j}+\varepsilon_{i j}\right).
\end{equation}
The perturbation $\varepsilon_{i j}$, can be expanded in terms of
hyperspherical harmonics on ${\bf S}^3$
\begin{equation}
\varepsilon_{i j}=\sum_{n} a_{n} \Omega_{i j} Q_{n}+b_{n} L_{i j
n}+c_{n} O_{i j n}+d_{n} U_{i j n}.
\end{equation}
The index $n$ actually represents three indices, but we have omitted
them for notational simplicity. The $Q_n$ are the scalar harmonics
on the 3-sphere. The $L_{ijn}$ are given in terms of $Q_n$ and
$\Omega_{i j}$. The $O_{ijn}$ are defined in terms of the transverse
vector harmonics, and the $U_{ijn}$ are the transverse traceless
tensor harmonics.

The matter field is represented by a conformally invariant scalar
field $\phi$, which can be expanded in terms of hyperspherical
harmonics $Q_n$ on ${\bf S}^3$, namely
\begin{equation}\label{expansionhyper}
\phi(\tau,x^i)= a^{-1}(\tau) \sum_{nlm} \phi_{nlm}(\tau)
Q_{nlm}(x^i),
\end{equation}
where $\phi_{nlm}$  are the coefficients of the scalar harmonics,
and $n=1,2,3, \ldots$; $l=0,1, \ldots, n-1$, $m=-l,-l+1, \ldots,l$.
From now on $n$ will represent the labels $n,l,m$.

In a suitable gauge, the coefficients, $a_n$, $b_n$, and $c_n$ can
be set to zero, and considering the case without gravitons, we can
also make $d_n=0$. Choosing that gauge, we write the 3-metric as
\begin{equation}
h_{i j}=a^{2} \, \Omega_{i j}.
\end{equation}
The wave function, $\Psi$ is then a function of the scale factor $a$
and of the coefficients of the scalar harmonics $\phi_n$.

As a result, the WDW equation for the wormhole is just the sum of a
collection of harmonic oscillators for the matter field modes, minus
a harmonic oscillator in the radius $a$ of the 3-sphere ${\bf S}^3$
\cite{Hawking:1990jb}
\begin{equation}\label{WDworm}
\bigg[\sum_n \bigg( -{\partial^2 \over \partial \phi^2_n} + n^2
\phi_n^2\bigg)-\bigg( -{\partial^2 \over \partial \phi^2_n} +
a^2\bigg)\bigg]\Psi(a,\phi_n)=0.
\end{equation}
This equation expresses the fact that the total energy of the
wormhole is zero because the positive energy of the matter field is
balanced by the gravitational energy.

Note that upon quantization, the canonical momenta take the form
\begin{align}
&P_{\phi_{n}} \to -\im \frac{\partial}{\partial \phi_{n}}, \\ \notag
&P_{a} \to -\im \frac{\partial}{\partial a}.
\end{align}
The quantum state of the closed Universe is then given by $\Psi =
\Psi(a,\phi_n).$

The matter modes $\phi_n$ do not interact with $a$, and the solution
to \eqref{WDworm} will be a product of a wave function related to
the gravitational part (a function of the radius $a$) times a wave
function which is a product of functions of the modes $\phi_n$. That
is
\begin{equation}
\Psi=\psi_m(a)\prod_{n}\psi_n(\phi_n).
\end{equation}
If we want to get solutions of the WDW equation \eqref{WDworm} that
represent wormholes then we need to consider the following boundary
conditions:
\noindent
$({\rm a})$$\Psi$ should be exponentially
damped at large values of the radius $a$.

\noindent
$({\rm b})$ $\Psi$ should be regular at $a=0$.

This is because $\Psi$ should represent an asymptotically Euclidean
region for large $a$ ($a \to \infty$), and there should be no
singularities as $a \to 0$. Thus, the wave function $\Psi$ will be
the product of a harmonic oscillator wave function in $a$ times the
harmonic oscillator wave functions in the matter fields.
\begin{equation}
\Psi(a,\phi_n)=H_m(a)e^{-a^2/2} \times \prod_n H_{m_n}(\phi_n
\sqrt{n}) e^{-n\phi_n^2/2}.
\end{equation}
Hence, we have a discrete spectrum of wormholes. In other words, in
the $n$th level of the harmonic oscillator, we have $m$ scalar
particles.

Note that solutions of the WDW equation are independent of the lapse
function, that is, the WDW equation is the same in the Lorentzian
and Euclidean regime. Then, how do we know if we are talking about a
Lorenzian or a Euclidean solution? The answer relies in the boundary
conditions. If the wave function is oscillatory, we have a Friedmann
Universe, but if we have an exponentially damped wave function then
we have an Euclidean wormhole.

\section{Quantum Euclidean wormholes in Ho$\check{\rm r}$ava-Lifshitz gravity}
\label{sec:QEWHLG}

Inspired by Hawking's treatment of quantum wormholes, the aim is to
find the WDW equation for the model of HL gravity coupled to a
non-relativistic scalar field, which we also consider as a
perturbation, that is, expanded in terms of hyperspherical harmonics
on ${\bf S}^3$.

Quantum cosmology in Ho$\check{\rm r}$ava-Lifshitz gravity
considering the general FLRW metric
\begin{equation}\label{flrwclosed}
ds^2 = -N^2(t) dt^2 + a^2(t)\left(\frac{dr^2}{1-k r^2} +
r^2(d\theta^2 +\sin^2 \theta d\phi^2)\right),
\end{equation}
where $k=1,0,-1$ for closed, flat or open Universe, respectively,
has been extensively studied (see for instance,
\cite{Bertolami:2011,Vakili:2013wc,Cordero:2017egl}).

In this background the HL action is given by
\begin{align}\label{puregravityhl}
S_{\text{HL}}=&\frac{1}{2}\int d t \left(\frac{N}{a}\right)\left[-(3
\lambda-1)\left(\frac{a \dot{a}}{N}\right)^2+2a^2-\frac{2 \Lambda
a^{4}}{3}-g_{\text{r}}-\frac{g_{\text{s}}}{a^{2}}\right],
\end{align}
where
\begin{align}
g_{\text{r}}=24\pi^2\left(3 g_{2}+g_{3}\right), \quad
g_{\text{s}}=288 \pi^4\left(9 g_{4}+3 g_{5}+g_{6}\right).
\end{align}
Note that the spatial integration over the 3-sphere, $\int_{{\bf
S}^3} \sqrt{h} \, d^3x=2 \pi^2$, has already been performed.

Now, for the scalar field action, we treat the scalar field as a
perturbation and expand it in terms of hyperspherical harmonics on
${\bf S}^3$
\begin{equation}\label{expansionagain}
\phi(t,x^i)= a^{-1}(t) \sum_{n} \phi_{n}(t) Q_{n}(x^i),
\end{equation}
where $n=1,2, \dots$.

The scalar harmonics are eigenfunctions of the Laplace-Beltrami
operator $\Delta$ associated to the metric $h_{ij}$, that is
\cite{Lindblom:2017maa}
\begin{equation}\label{eigen}
\Delta Q_{n}=-\frac{n(n+2)}{a^{2}} Q_{n}.
\end{equation}
They also obey the orthonormality condition
\begin{equation}
\frac{1}{a^3} \int d^3x \sqrt{h} \, Q_n Q_{n^{\prime}}^{*}=\delta_{n
n^{\prime}}.
\end{equation}
Plugging \eqref{expansionagain} into the scalar field action
\eqref{scalarHL}, we obtain the following result
\begin{equation}\label{purematterhl}
S_{\text{m}}=\frac{1}{2}\int dt \,
\left(\frac{N}{a}\right)\left[\frac{(3 \lambda-1)}{2}
\left(\frac{a\dot{\phi}_n}{N}\right)^2-\phi_{n}^2\left(\beta_1
+\frac{\beta_2}{a^2}+\frac{\beta_3}{a^4}\right)\right],
\end{equation}
with
\begin{equation}
\beta_1=c_1n(n+2), \quad \beta_2=c_2n^2(n+2)^2, \quad
\beta_3=c_3n^3(n+2)^3.
\end{equation}
Now, taking \eqref{puregravityhl} and \eqref{purematterhl}, we write
the total action as
\begin{align}
S=&\frac{1}{2}\int dt \left(\frac{N}{a}\right)\left[-(3 \lambda-1)\left(\frac{a \dot{a}}{N}\right)^2+2a^2-\frac{2 \Lambda a^{4}}{3}-g_{\text{r}}-\frac{g_{\text{s}}}{a^{2}}\right] \notag \\
&+\frac{1}{2} \sum_n \int dt \,
\left(\frac{N}{a}\right)\left[\frac{(3 \lambda-1)}{2}
\left(\frac{a\dot{\phi}_n}{N}\right)^2-\phi_{n}^2\left(\beta_1
+\frac{\beta_2}{a^2}+\frac{\beta_3}{a^4}\right)\right].
\end{align}
We compute the full Hamiltonian by means of the Legendre
transformation
\begin{equation}
H=\dot{a}P_a+\dot{\phi}_n P_{\phi_{n}}-\mathcal{L},
\end{equation}
where a sum over $n$ is implied, ${\cal L}$ is the Lagrangian of the
total action $S$ and the canonical conjugate momenta are given by
\begin{equation}
P_{a}=\frac{\partial \mathcal{L}}{\partial \dot{a}}=- \gamma\frac{a
\dot{a}}{N}, \quad P_{\phi_{n}}=\frac{\partial \mathcal{L}}{\partial
\dot{\phi}_n}=\gamma\frac{a \dot{\phi}_n}{2N},
\end{equation}
where $\gamma=(3\lambda-1)$.

The total Hamiltonian is then
\begin{align}
H=\left(\frac{N}{a}\right)\bigg[-\frac{1}{2}\frac{P_{a}^2}{\gamma}+\frac{P_{\phi_{n}}^2}{\gamma}-\frac{1}{2}\left(2a^2+\frac{2
\Lambda a^{4}}{3}+g_{\text{r}}+\frac{g_{\text{s}}}{a^{2}} \right)
+\frac{1}{2}\phi_{n}^2\left(\beta_1
+\frac{\beta_2}{a^2}+\frac{\beta_3}{a^4}\right)\bigg].
\end{align}
To obtain the WDW equation for this model, we promote the
Hamiltonian to an operator acting on the wave function of the
Universe, $\Psi(a,\phi_{n})$. We have to take into account
ambiguities in the operator ordering, thus, we write
\begin{equation}
\widehat{P}_{a}^{2} \mapsto -\frac{1}{a^{p}}
\frac{\partial}{\partial a}\left(a^{p} \frac{\partial}{\partial
a}\right), \quad \widehat{P}_{\phi_{n}}^{2} \mapsto - {\partial^2
\over \partial \phi^2_n},
\end{equation}
where the operator ordering ambiguity between the operators $a$ and
$P_a$ is reflected in the arbitrary constant $p$, and becomes
important only for very small values of the scale factor $a$. In
quantum cosmology there exist two popular choices of this parameter.
If $p=1$ we have the so called {\it Laplace-Beltrami operator
ordering}, but if $p=-1$ then we have the {\it Vilenkin ordering}
\cite{Steigl:2005fk}.

Finally, the WDW equation for a non-relativistic scalar field
coupled to HL gravity is given by
\begin{align} \label{wdw-hl}
& \bigg \{{\partial^2 \over \partial a^2}+\frac{p}{a}{\partial \over \partial a}+\gamma\left(-2a^2+\frac{2 \Lambda a^{4}}{3}+g_{\text{r}}+\frac{g_{\text{s}}}{a^{2}}\right) \notag\\
&+2 \sum_n  \left[-{\partial^2 \over \partial
\phi^2_n}+\frac{1}{2}\gamma \phi_{n}^2\left(\beta_1
+\frac{\beta_2}{a^2}+\frac{\beta_3}{a^4}\right)\right]\bigg \}
\Psi(a,\phi_{n})=0.
\end{align}
The objective is to try to find solutions (at least for certain
limits) to the equation \eqref{wdw-hl} that are consistent with the
Hawking and Page proposal for Euclidean quantum wormholes. In other
words, such solutions have to satisfy the following boundary
conditions:

\noindent
(a). $\Psi$ should decay exponentially as the radius $a
\to \infty$.

\noindent
(b). $\Psi$ should be regular as $a \to 0$.

\subsection{Solution to the WDW equation in the limit $a \to \infty$}

The large $a$ limit corresponds to the very late Universe, which is
dominated by the curvature and the cosmological constant. For large
values of the scale factor, the term with the parameter $p$, and the
terms with constants $g_{\text{s}}$, $\beta_2$, and $\beta_3$, all
go to zero. We can also neglect the constant $g_{\text{r}}$ because
it is very small compared to the surviving terms. The curvature and
cosmological constant terms are the only ones that dominate in this
limit. Hence, equation \eqref{wdw-hl} becomes
\begin{align} \label{wdw-hl1}
\bigg \{{\partial^2 \over \partial a^2}+\gamma\left(-2a^2+\frac{2
\Lambda a^{4}}{3}\right) +2 \sum_n  \left[-{\partial^2 \over
\partial \phi^2_n}+\frac{1}{2}\gamma \beta_1 \phi_n^2 \right]\bigg
\} \Psi=0.
\end{align}
For $\Lambda=0$ we obtain
\begin{align} \label{wdw-hl2}
\bigg[{\partial^2 \over \partial a^2}-\omega_0^2a^2 +2 \sum_n
\left(-{\partial^2 \over \partial
\phi^2_n}+\omega_1^2\phi_n^2\right)\bigg] \Psi=0,
\end{align}
where $\omega_0^2=2\gamma$, and $\omega_1^2=\frac{1}{2}\gamma
\beta_1$.

The equation \eqref{wdw-hl2} resembles the WDW equation
\eqref{WDworm} obtained by Hawking. As we expected, we are back to
the usual GR quantum cosmology. Actually, we are in the IR fixed
point where $\lambda=1$ and as we mentioned in Section 2, this limit
should be stable only non-perturbatively. The difference appears in
the level of the Friedmann equation, but at the level of the
Hamiltonian constraint this problem is not evident.

We note that the total wave function is separable, and it is just a
product of a gravitational part of the wave function times a wave
function for the matter fields
\begin{equation}
\Psi(a,\phi_n)=\psi(a)\varphi(\phi_n).
\end{equation}
Thus, the WDW equation separates into\footnote{We have taken the
separation constant equal to zero.} the following equations
\begin{equation}\label{harmonicgrav}
\left( -{\partial^2 \over \partial
a^2}+\omega_0^2a^2\right)\psi(a)=0,
\end{equation}
\begin{equation}\label{harmonicmatter}
\sum_n  \left(-{\partial^2 \over \partial
\phi^2_n}+\omega_1^2\phi_n^2\right)\varphi(\phi_n)=0.
\end{equation}
Note that the coefficients $\phi_n$, appear in the WDW equation like
the coordinate $x$ of a harmonic oscillator with frequency
$\omega_1^2$ independent of $a$. Indeed, we have two harmonic
oscillator--like equations, one for $a$, and one for $\phi_n$. The
solution of equation \eqref{harmonicgrav} is then
\begin{equation}
\psi_E(a)=N_{E} \,
\exp\bigg(\frac{-a^2\omega_0}{2}\bigg)H_E(a\sqrt{\omega_0}),
\end{equation}
where $N_E=\left(\frac{\omega_0}{\pi \,
2^{2E}(E!)^2}\right)^\frac{1}{4}$ is the normalization constant, and
$H_E$ are the Hermite polynomials, with $E=0,1,2, \dots$

Similarly, the solution of equation \eqref{harmonicmatter} is given
by
\begin{equation}\label{solharmonicmatter}
\varphi_m(\phi_n)=N_{m} \prod_n \,
\exp\bigg(\frac{-\phi_n^2\omega_1}{2}\bigg)H_{m_n}(\phi_n\sqrt{\omega_1}),
\end{equation}
where $N_m=\left(\frac{\omega_1}{\pi \,
2^{2m}(m!)^2}\right)^\frac{1}{4}$ with $m=0,1,2, \dots$. These
solutions can be interpreted as corresponding to the closed Universe
containing $m$ scalar particles in the $n$th harmonic mode.

\begin{figure}[h]
\centering
\includegraphics[width=0.9\textwidth]{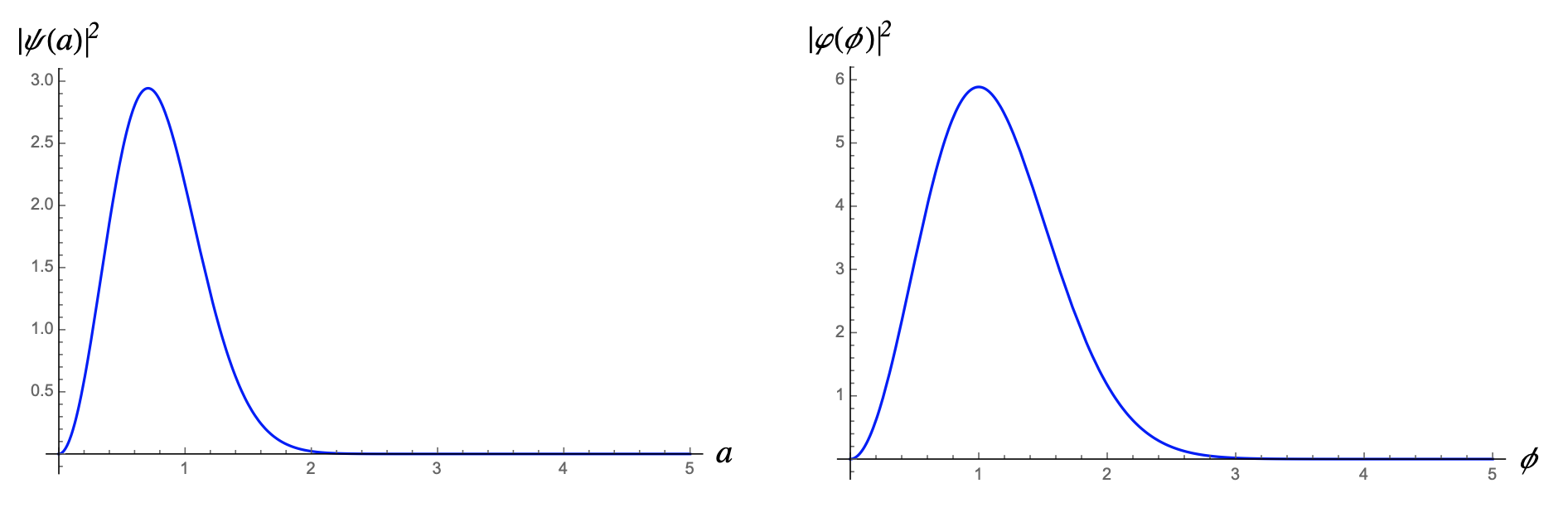}
\caption{On the left hand side is the plot of $|\psi(a)|^2$ for
$E=1$, and on the right hand side is the plot of $|\varphi(\phi)|^2$
for $m=1$ and $n=1$. We set $\lambda=1$, which is the value of the
parameter in the GR limit.} \label{hermiteclass}
\end{figure}

As seen in Figure \ref{hermiteclass}, both the gravitational sector
of the wave function and the matter wave function have an
exponentially asymptotic behavior:
\begin{equation}
\psi(a) \sim e^{-\frac{a^2}{2}\omega_0}, \quad \varphi(\phi_n) \sim
e^{-\frac{\phi_n^2   }{2}\omega_1}.
\end{equation}
The part of the matter fields was plotted taking one scalar particle
($m=1$) in the first harmonic mode ($n=1$).

Therefore, the total wave function,
$\Psi(a,\phi_n)=\psi(a)\varphi(\phi_n)$, in the limit of large $a$,
agrees with the boundary condition (a) of the Hawking-Page
conjecture.

\subsection{Solution to the WDW equation in the limit $a \to 0$}

Now we are interested in studying the case when the scale factor is
small, or $a \to 0$. This means that we are in the early times of
the cosmic evolution. Precisely for short distances HL theory is
more appropriated than GR since it is well behaved in this limit and
it should give sensible results near the singularity. In this
scenario, the surviving terms in the gravitational part of equation
\eqref{wdw-hl} are the ones with HL parameters $g_{\text{r}}$ and
$g_{\text{s}}$, as well as the one with the parameter $p$. When $a$
is small, we cannot neglect the quantum effects, which means that
the operator ordering parameter $p$ becomes more significant.

In the matter fields part of equation \eqref{wdw-hl}, we consider
that $\frac{1}{a^2}$ is smaller than $\frac{1}{a^4}$ when $a \to 0$,
so we can neglect the former one. Therefore, in this limit, the
higher-order terms dominate, and the WDW equation \eqref{wdw-hl}
reads
\begin{equation}
\label{smallaeq}
\bigg \{ {\partial^2 \over \partial a^2}+\frac{p}{a}{\partial \over
\partial a} +\gamma\left(g_{\text{r}}+\frac{g_{\text{s}}}{a^2}\right) +2 \sum_n
\left[-{\partial^2 \over \partial \phi^2_n}+\frac{1}{2}
\frac{\gamma\beta_3}{a^4} \phi_n^2\right]\bigg \} \Psi=0.
\end{equation}
In order to obtain solutions for this equation, we need to make the
following change of variables for each $\phi_n$
\cite{SolomonsPhDthesis1993}
\begin{equation}\label{changecoord}
\eta_n =\frac{\phi_n^2}{2a^2}.
\end{equation}
With the change of coordinates $(a,\phi_n) \to (a,\eta_n)$, the WDW
equation becomes
\begin{equation}
\left[ a^2 {\partial^2 \over \partial a^2}+ pa{\partial \over
\partial a}+\gamma\left(a^2 g_{\text{r}}+g_{\text{s}}\right) -2 \sum_n
\left(2\eta_n {\partial^2 \over \partial \eta^2_n} + {\partial \over
\partial \eta_n} -\gamma \beta_3 \eta_n \right)\right]\Psi=0.
\end{equation}
Hence, the change of coordinates introduced earlier makes the wave
function separable:
\begin{equation}
\Psi(a,\eta_n)=\psi(a)\varphi(\eta_n).
\end{equation}
The WDW equation \eqref{smallaeq} separates into
\begin{equation}\label{besselhl}
\left(a^2 {d^2 \over da^2}+pa {d \over da}+\gamma\left(a^2
g_{\text{r}}+g_{\text{s}}\right) + \chi \right)\psi(a)=0,
\end{equation}
and
\begin{equation}\label{hermitehl}
\sum_n \left(2\eta_n {d^2 \over d\eta^2_n} + {d \over d\eta_n} -
\gamma \beta_3 \eta_n + \chi \right)\varphi(\eta_n)=0.
\end{equation}
The general solution of equation \eqref{besselhl} is given by a
linear combination of the Bessel functions and Neumann functions
\cite{bellBook2004}
\begin{equation}\label{generalsolgrav}
\psi(a)=C_1 \, a^{\frac{1-p}{2}} \, J_{\nu}(a\sqrt{\gamma
g_{\text{r}}})+C_2 \, a^{\frac{1-p}{2}} Y_{\nu}(a\sqrt{\gamma
g_{\text{r}}}),
\end{equation}
where the order of the Bessel function is $\nu = \frac{1}{2}
\sqrt{1+p(p-2)-4(\gamma g_{\text{s}}+\chi)}$, which depends on the
value of the parameters $\lambda$ and $p$. As previously mentioned,
$\lambda$ is a dynamical coupling constant which can take different
values in the UV regime. Here, we will work with values $\lambda >
1/3$ such that $(3\lambda-1) \neq 0$. Also, in order to avoid a
complex argument in the Bessel functions, we will take $g_{\text{r}}
> 0$. The {\it bounds} for $p$ and $g_{\text{s}}$ will be given
below.

Since we are considering the limit $a \to 0$, not both functions in
the general solution \eqref{generalsolgrav} are admissible. This can
be seen from the asymptotic forms of the Bessel and Neumann
functions
\begin{equation}
J_{\nu}(z) \sim z^{\nu}, \quad Y_\nu(z) \sim -z^{-\nu}.
\end{equation}
Notice that the Neumann functions, $Y_\nu(z)$, diverge at the
origin, so we set $C_2=0$, and keep only the asymptotic form of
$J_\nu(z)$ for small $z$.

\begin{figure}[h]
\centering
\includegraphics[width=0.9\textwidth]{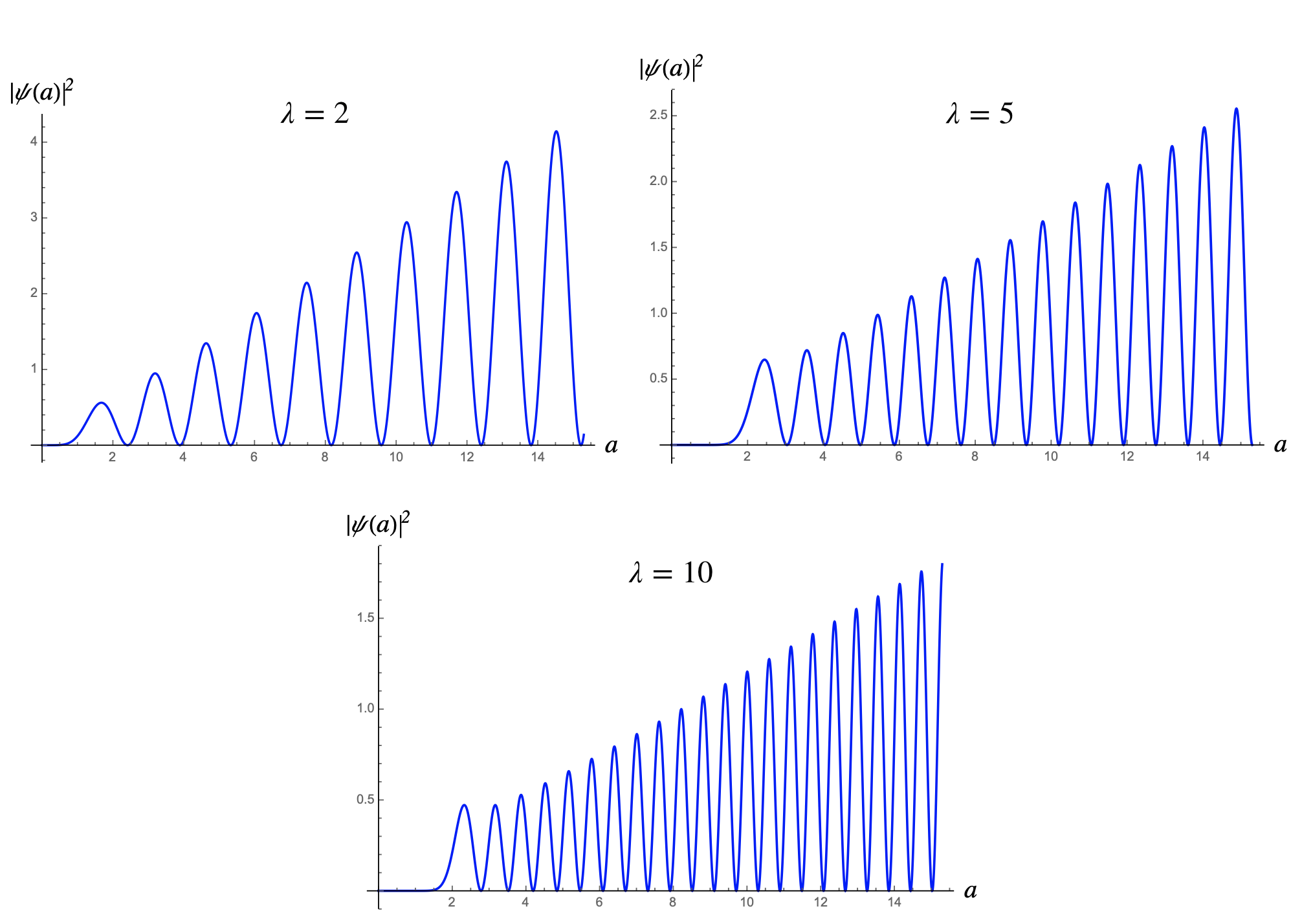}
\caption{Plot of the square of the wave function, $\psi(a) \simeq
a^{\frac{1-p}{2}} \, J_{\nu}(a\sqrt{\gamma g_{\text{r}}})$, for
$p=-1$, $g_{\text{r}}=1$, $g_{\text{s}}=-1$, and for three different
values of the parameter $\lambda$.} \label{bessel}
\end{figure}

Therefore, for small $a$ the general solution \eqref{generalsolgrav}
reads
\begin{equation}
\psi(a) \sim \, a^{\frac{1-p}{2}+\nu}.
\end{equation}
This solution is indeed regular at $a=0$ satisfying the Hawking-Page
boundary condition (b). However, one needs to be careful with the
choice of operator ordering because for $p>1$ we have a divergence.
Nevertheless, for $p < 1$ the divergence is avoided, and the
wormhole boundary condition is satisfied. As for $g_{\text{s}}$, we
can restrict their value by demanding $\nu$ to be real. Hence, the
bound for $g_{\text{s}}$ is
\begin{equation}
g_{\text{s}} < \frac{1+p(p-2)}{4 \gamma}-\frac{\chi}{\gamma}.
\end{equation}
In Figure \ref{bessel}, we show the plot of the solution of the
gravitational part \eqref{generalsolgrav} for $C_2=0$. We choose
three different values of $\lambda$. The plot \ref{bessel} shows
that the higher the value of the parameter $\lambda$, the higher the
frequency of oscillation of $\psi(a)$ for $a$ far away from the
origin. It is also observed that the amplitude decreases as
$\lambda$ grows.  This does not contradict the Hawking-Page boundary
conditions because we are interested in the behavior of the wave
function for small $a$, and in that region, the solution is indeed
regular, i.e. $\psi(a)=0$ for $a=0$.
We still need to give the allowed values of $\chi$. This will arise
naturally as we explore the solution of equation \eqref{hermitehl}.

Equation \eqref{hermitehl} is a Hermite-like equation, hence its
solution can be written as
\begin{equation}\label{solmatterhermite}
\varphi_m(\eta_n)=A_m \prod_n
e^{-\frac{\rho}{2}\eta_n}H_{m_n}(\sqrt{\rho \eta_n}),
\end{equation}
where $\chi=\rho(m+\frac{1}{2})$ for $m=0,1,2,\dots$, $\rho^2=\gamma
\beta_3$, and $A_m=(2^m m!)^{-\frac{1}{2}}$ is the normalization
constant.

Asymptotically, the solution behaves like
\begin{equation}
\varphi(\eta) \sim e^{-\frac{\rho}{2}\eta},
\end{equation}
which implies that as $\eta \to \infty$ ($a \to 0$) the solution is
regular.

\begin{figure}[h]
\centering
\includegraphics[width=0.9\textwidth]{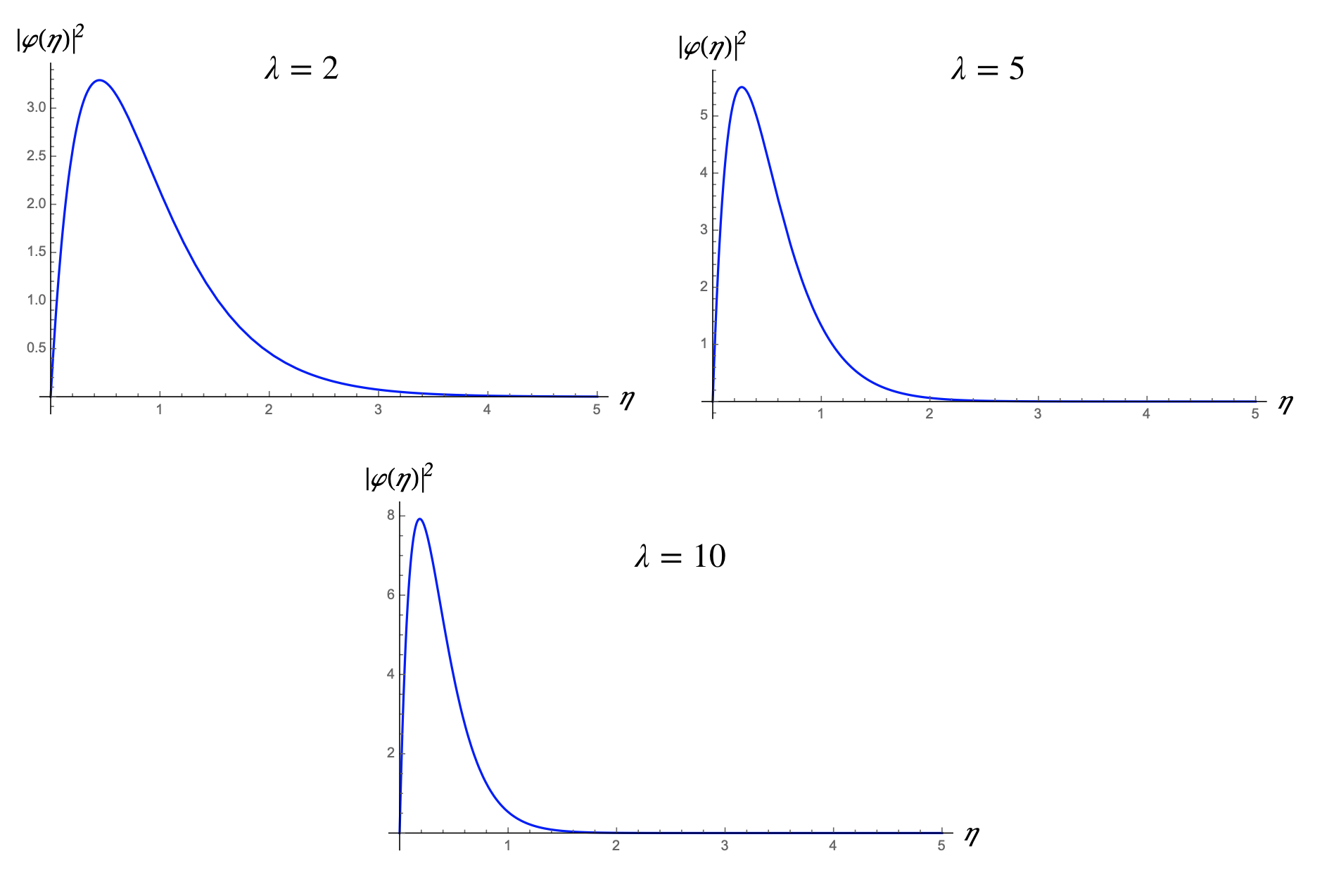}
\caption{Plot of the square of the wave function, $\varphi_1(\eta)
\simeq e^{-\frac{\rho}{2}\eta}H_{1}(\sqrt{\rho \eta})$, for $n=1$,
$m=1$ and for three different values of the parameter $\lambda$.}
\label{hermite}
\end{figure}

In Figure \ref{hermite}, the solution \eqref{solmatterhermite} is
plotted considering one scalar particle ($m=1$) in the first excited
state ($n=1$) for different values of the parameter $\lambda$. As we
can see, for higher values of the parameter $\lambda$, the peak of
the distribution becomes narrower.

Finally, the full wave function in this limit behaves like
\begin{equation}\label{hlsolutionhp}
\Psi(a,\eta) \sim a^{\frac{1-p}{2}+\nu} \, e^{-\frac{\rho}{2}\eta}.
\end{equation}
which is of course regular near the origin, obeying the Hawking-Page
boundary condition (b).

It is worth mentioning that this solution resembles the one found
when a massive scalar field is minimally coupled to gravity
\cite{SolomonsPhDthesis1993}.

The solutions here presented were obtained for zero cosmological
constant, but they can be extended for $\Lambda \neq0$.

\subsection{Solution to WDW equation for $\Lambda\neq0$}
Solutions to the WDW equation in the context of a FLRW
minisuperspace model have been studied in \cite{Bertolami:2011}
considering all the possible values of the cosmological constant.
Here we will analyze some of the results in the context of wormhole
solutions.

\noindent
$\bullet$ Solution for $\Lambda>0$

In the limit $a \to \infty$, the term with $a^2$ is smaller compared
to the term with $a^4$, so we only keep the latter. Also, the
operator ordering parameter $p$ is no longer significant, and we end
up with the following equations:
\begin{equation}\label{oscillatory}
\left({\partial^2 \over \partial a^2}+ g_\Lambda a^4\right)\psi(a)=0
\end{equation}
and
\begin{equation}\label{harmonicmatter2}
\sum_n  \left(-{\partial^2 \over \partial
\phi^2_n}+\omega_1^2\phi_n^2\right)\varphi(\phi_n)=0,
\end{equation}
where $g_\Lambda=\frac{2\gamma\Lambda}{3}$, and the separation
constant has been set to zero so that we can obtain physically
meaningful solutions.

The solution to Eq. \eqref{oscillatory} is given by a combination of
Bessel and Neumann functions
\begin{equation}\label{soloscillatory}
\psi(a)=\overline{C}_{1} \sqrt{a} \, J_{1 /
6}\left(\frac{\sqrt{g_\Lambda}}{3} a^{3}\right)+\overline{C}_{2}
\sqrt{a} \, Y_{1 / 6}\left(\frac{\sqrt{g_\Lambda}}{3} a^{3}\right),
\end{equation}
while the solution of \eqref{harmonicmatter2} is given in terms of
Hermite functions, and it is the same as in
\eqref{solharmonicmatter}.

The behavior of \eqref{soloscillatory} for large arguments is
oscillatory. This can be easily seen by making use of the asymptotic
expansion for Bessel and Neumann functions for $|z| \rightarrow
\infty$
\begin{align}
&J_{\nu}(z) \sim \sqrt{\frac{2}{\pi z}} \cos \left(z-\frac{\nu \pi}{2}-\frac{\pi}{4}\right), \notag \\
&N_{\nu}(z) \sim \sqrt{\frac{2}{\pi z}} \sin \left(z-\frac{\nu
\pi}{2}-\frac{\pi}{4}\right).
\end{align}
Hence, using these expressions, the asymptotic behavior of $\psi(a)$
for large $a$ reads
\begin{equation}
\psi(a) \sim \frac{C_{1}}{a} \cos \left(\frac{\sqrt{g_\Lambda}}{3}
a^{3}-\frac{\pi}{12}-\frac{\pi}{4}\right) +\frac{C_{2}}{a} \sin
\left(\frac{\sqrt{g_\Lambda}}{3}
a^{3}-\frac{\pi}{12}-\frac{\pi}{4}\right),
\end{equation}
where $C_{i}=\overline{C}_{i} \sqrt{6 / \pi \sqrt{g_{\Lambda}}}$,
for $i=1,2$.

\begin{figure}[h]
\centering
\includegraphics[width=0.5\textwidth]{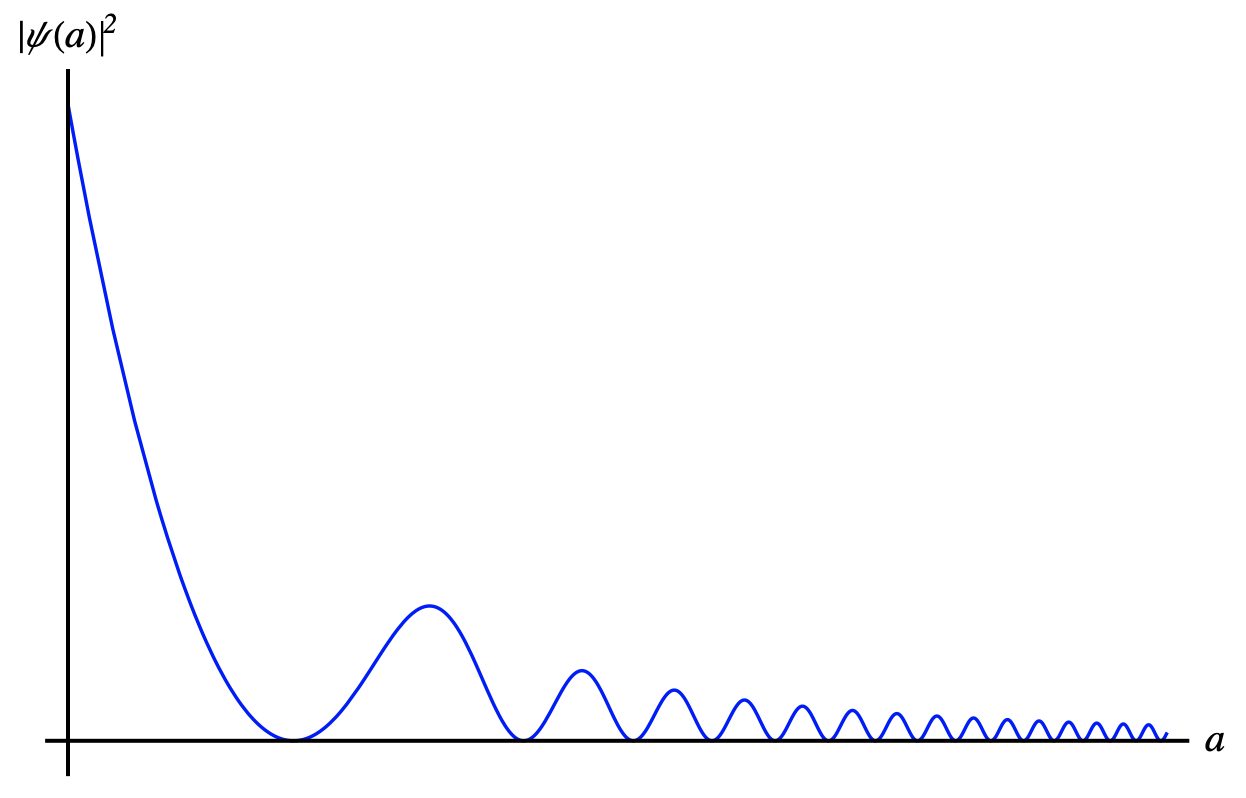}
\caption{Qualitative plot of $|\psi(a)|^2$ for $\lambda=1$, which is
the value that the parameter takes in the GR limit. The wave
function $\psi(a)$ oscillates for large $a$.} \label{besselJY}
\end{figure}

As we can see in Figure \ref{besselJY}, $\psi(a)$ has an oscillatory
behavior for large $a$. Even if we take into account the solution of
the matter part (which goes like $\exp(-\phi^2)$), the oscillatory
behavior will not be suppressed. Hence, the total wave function
$\Psi(a,\phi)$, for large $a$, does not satisfy the required
boundary condition, and for that reason it does not describe a
quantum wormhole. However, we may interpret this wave function as a
Lorentzian or Friedmann Universe.

In the limit $a \to 0$ we are back in the HL regime, where the
cosmological constant is no longer significant. Indeed, the
solutions in that limit will behave like in equation
\eqref{hlsolutionhp}, and satisfy the wormhole boundary condition
(b).

From the above analysis, we see that the presence of a positive
cosmological constant makes the wave function of the Universe to
have an oscillatory behavior in the limit $a \to \infty$. For that
reason, the Hawking-Page wormholes cannot exist in this scenario.

\vskip .5truecm

\noindent
$\bullet$ Solution for $\Lambda<0$

For $a \to \infty$, and $g_\Lambda <0$ the equations to solve are
\begin{equation}\label{oscillatory2}
\left({\partial^2 \over \partial a^2}-( -g_\Lambda)
a^4\right)\psi(a)=0,
\end{equation}
and
\begin{equation}\label{harmonicmatter3}
\sum_n  \left(-{\partial^2 \over \partial
\phi^2_n}+\omega_1^2\phi_n^2\right)\varphi(\phi_n)=0.
\end{equation}
The solution of the gravitational part \eqref{oscillatory2} is a
combination of the modified Bessel functions, that is
\begin{equation}\label{soloscillatory2}
\psi(a)=C_{1} \sqrt{a} \, I_{1 / 6}\left(\frac{\sqrt{-g_\Lambda}}{3}
a^{3}\right)+C_{2} \sqrt{a} \, K_{1 /
6}\left(\frac{\sqrt{-g_\Lambda}}{3} a^{3}\right),
\end{equation}
while the solution of \eqref{harmonicmatter3} is again given in
terms of the Hermite functions, and it is the same as in
\eqref{solharmonicmatter}.

Now, let us write the asymptotic form of the modified Bessel
functions for large argument ($|z| \to \infty$)
\begin{equation}
I_{\nu}(z) \sim \frac{e^{z}}{\sqrt{2 \pi z}}, \quad K_{\nu}(z) \sim
\sqrt{\frac{\pi}{2 z}} e^{-z}.
\end{equation}
From these expressions we notice that $I_{\nu}(z)$ grows
exponentially as $z \to \infty$, hence we only keep $K_{\nu}(z)$
because it decays exponentially for large $z$ and satisfies the
wormhole boundary condition (a).

\begin{figure}[h]
\centering
\includegraphics[width=0.5\textwidth]{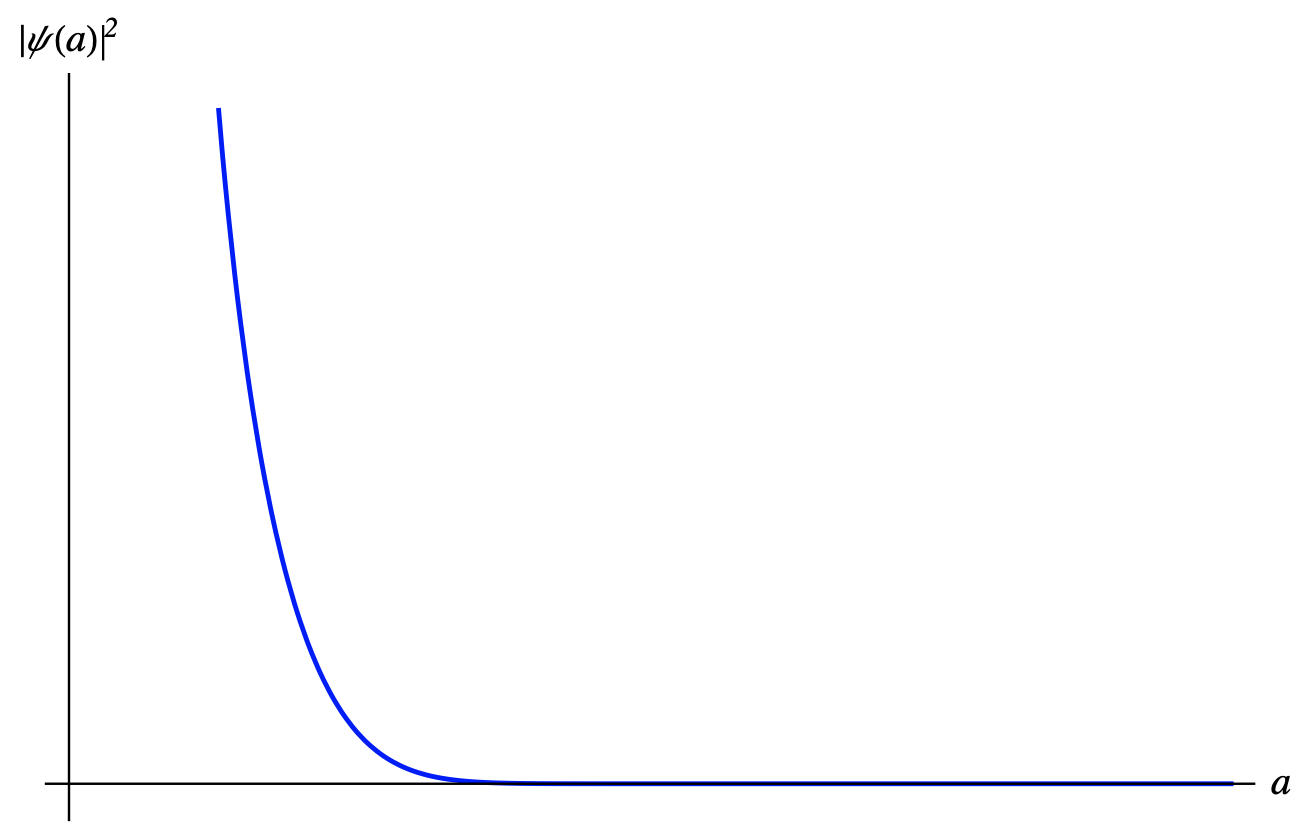}
\caption{Qualitative plot of $|\psi(a)|^2$ for $\lambda=1$. The wave
function $\psi(a)$ is exponentially damped for large $a$, but it is
also highly suppressed.} \label{modbessel}
\end{figure}

Therefore, the asymptotic solution of \eqref{oscillatory2} is
\begin{equation}
\psi(a) \sim \frac{1}{a} \, e^{(-\sqrt{-g_{\Lambda}} / 3) a^{3}}.
\end{equation}
The plot of this solution is shown in Figure \ref{modbessel}. We see
that for large $a$, the wave function is exponentially damped,
however it is highly suppressed due to the factor $e^{-a^3}$. The
full wave function in this limit is
\begin{equation}
\Psi(a,\phi) \sim \frac{1}{a} \, e^{(-\sqrt{-g_{\Lambda}} / 3)
a^{3}} e^{-\frac{\phi^2}{2}\omega_1}.
\end{equation}
It seems that in the limit of large $a$, we have a exponentially
damped wave function. Indeed, this solution satisfies the
Hawking-Page boundary condition (a).

On the other hand, in the region of the early Universe, that is, in
the limit $a \to 0$, the HL terms are the ones that dominate, and
the cosmological constant does not appear. Hence, we are back to the
case with solution \eqref{hlsolutionhp}, which agrees with the
wormhole boundary condition (b).

The performed analysis shows that for a negative cosmological
constant we have a exponentially (but highly suppressed) damped
behavior in the limit $a \to \infty$, and a regular behavior in the
limit $a \to 0$. This satisfies both Hawking-Page boundary
conditions. Hence, in a Universe with negative cosmological
constant, wormhole configurations may exist.

\section{Final remarks}
\label{sec:final}

We have studied Euclidean quantum wormholes in the context of the
projectable version of the HL gravity. We considered a FLRW closed
Universe minisuperspace model coupled to a scalar field, which we
treated as a perturbation, which is expanding on hyperspherical
harmonics on the 3-sphere. We obtained a WDW equation by canonically
quantizing such model. As expected, the gravitational part of the
WDW equation is fully characterized by the scale factor. On the
other hand, in the matter part of the equation, the coefficients of
the scalar harmonics and the scale factor appear. This is a
consequence of the higher-order spatial derivative terms in the
matter action. In order to obtain analytical solutions to the full
WDW equation (gravity$+$matter) we solved it by considering limiting
cases of small and large scale factor.

In the limit of the very late Universe (large scale factor) and
considering a vanishing cosmological constant, the resulting WDW
equation turned out to be a sum of two harmonic oscillator
equations, one for the matter fields and one for the scale factor.
The WDW equation is similar to the one found in GR when considering
a model of gravity coupled to a conformally invariant scalar field
as shown by Hawking \cite{Hawking:1990jb}. In this case, the
solution of the WDW equation is a product of two harmonic oscillator
wave functions, which are given in terms of Hermite polynomials.
This solution is exponentially damped at large values of the radius
$a$ agreeing with the Hawking-Page boundary condition (a). It is
also important to point out that this case corresponds to the
Hamiltonian constraint in GR, where the additional scalar degree of
freedom of the projectable theory is not evident.

In the limit of the very early Universe (small scale factor) things
get more interesting because the quantum effects cannot be
neglected, and the operator ordering parameter $p$ becomes more
significant. In this regime, the HL terms dominate, and hence the
cosmological constant term can be neglected. The WDW equation turned
out to be non-separable because the term $\frac{\phi^2}{a^4}$
appeared in the matter part. However, we avoided this problem by
using a suitable choice of coordinates. Fortunately, the equations
obtained after this change have known solutions. Specifically, the
equation of the gravitational part is a Bessel-like equation, thus
their solutions are a linear combination of Bessel functions of the
first and second kind. We kept the function $J_\nu(z)$ because it is
the one that satisfies the wormhole boundary condition (b), that is,
it is regular in the limit of small values of the scale factor.
However, as the value of $a$ increases we start to see an
oscillatory behavior. This does not contradict the Hawking-Page
conjecture, it just means that $\psi(a)$ is regular in a different
way.

The matter equation is given in terms of the new coordinate
$\eta=\frac{\phi^2}{2a^2}$, and also has known solutions, namely,
Hermite polynomials. Notice that at small values of the radius $a$,
$\eta \to \infty$. In this limit, the solution behaves like
$\varphi(\eta) \sim e^{-\frac{\rho}{2}\eta}$. Therefore, it
eventually goes to zero, and hence is also regular, which agrees
with the Hawking-Page boundary condition (b).

The solution obtained in the limit of small radius $a$ is similar to
the one obtained when a massive scalar field is minimally coupled to
Einstein's gravity. This observation is interesting because it means
that the HL gravity terms naturally behave like some kind of
'effective mass'.

We also gave different values of the running parameter $\lambda$ of
HL theory in the limit $a \to 0$. The effect of $\lambda$ on
$\psi(a)$ was decrease the amplitude. It is also observed the
increment on the frequency of oscillation, but it remained regular
at the origin, which is the region we are interested in. On the
other hand, its effect on $\varphi(\eta)$ was to make the function
more localized or peaked.

Supposedly, a necessary condition for wormhole-like solutions to
appear is to take $\Lambda=0$. However, we expanded our analysis by
considering the case of nonzero cosmological constant, which
dominates in the IR limit. As expected, for $\Lambda
>0$ we found oscillatory solutions, indeed, this happens because we
are in a classically allowed region, i.e. the WDW potential $V(a)
<0$. Thus, this solution does not satisfy the wormhole boundary
condition (a) meaning that wormholes cannot exist in this scenario.
For $\Lambda <0$ we obtained a damped solution. This is also not
surprising because it just means that we are in a classically
forbidden region, $V(a)>0$. This solution combined with the one
found in the limit $a \to 0$ tell us that wormholes might exist in a
Universe with negative cosmological constant.

Finally we want to mention that stable Euclidean wormhole solutions
do exist in Anti-de Sitter spacetime via the holographic
correspondence \cite{Maldacena:2004rf,Betzios:2019rds}. It would be
interesting to find a relation between these issues and the result
presented in our paper. In addition for future work, it would be
interesting, to look for numerical solutions for the full equation
(\ref{wdw-hl}) and analyze in detail and interpret physically these
solutions.

 \vspace{1cm}
\centerline{\bf Acknowledgments} \vspace{.5cm} A.Vazquez would like
to thank CONACyT for a grant.


\break

\end{document}